\documentstyle[epsf]{mn}

\newif\ifAMStwofonts

\ifoldfss
  \ifCUPmtlplainloaded \else
    \NewTextAlphabet{textbfit} {cmbxti10} {}
    \NewTextAlphabet{textbfss} {cmssbx10} {}
    \NewMathAlphabet{mathbfit} {cmbxti10} {} 
    \NewMathAlphabet{mathbfss} {cmssbx10} {} 
  \fi
  \ifAMStwofonts
    \ifCUPmtlplainloaded \else
      \NewSymbolFont{upmath} {eurm10}
      \NewSymbolFont{AMSa} {msam10}
      \NewMathSymbol{\upi}     {0}{upmath}{19}
      \NewMathSymbol{\umu}     {0}{upmath}{16}
      \NewMathSymbol{\upartial}{0}{upmath}{40}
      \NewMathSymbol{\leqslant}{3}{AMSa}{36}
      \NewMathSymbol{\geqslant}{3}{AMSa}{3E}

      \let\leq=\leqslant 
       
    \fi
  \fi
\fi 

\ifnfssone
  \newmathalphabet{\mathit}
  \addtoversion{normal}{\mathit}{cmr}{m}{it}
  \addtoversion{bold}{\mathit}{cmr}{bx}{it}
  \newmathalphabet{\mathbfit} 
  \addtoversion{normal}{\mathbfit}{cmr}{bx}{it}
  \addtoversion{bold}{\mathbfit}{cmr}{bx}{it}
  \newmathalphabet{\mathbfss} 
  \addtoversion{normal}{\mathbfss}{cmss}{bx}{n}
  \addtoversion{bold}{\mathbfss}{cmss}{bx}{n}
  \ifAMStwofonts
    \ifCUPmtlplainloaded \else
      \UseAMStwoboldmath
      \makeatletter
      \new@mathgroup\upmath@group
      \define@mathgroup\mv@normal\upmath@group{eur}{m}{n}
      \define@mathgroup\mv@bold\upmath@group{eur}{b}{n}
      \edef\UPM{\hexnumber\upmath@group}
      \new@mathgroup\amsa@group
      \define@mathgroup\mv@normal\amsa@group{msa}{m}{n}
      \define@mathgroup\mv@bold\amsa@group{msa}{m}{n}
      \edef\AMSa{\hexnumber\amsa@group}
      \makeatother
      \mathchardef\upi="0\UPM19
      \mathchardef\umu="0\UPM16
      \mathchardef\upartial="0\UPM40
      \mathchardef\leqslant="3\AMSa36
      \mathchardef\geqslant="3\AMSa3E

      \let\leq=\leqslant 

    \fi
  \fi
\fi 

\ifnfsstwo
  \DeclareMathAlphabet{\mathbfit}{OT1}{cmr}{bx}{it}
  \SetMathAlphabet\mathbfit{bold}{OT1}{cmr}{bx}{it}
  \DeclareMathAlphabet{\mathbfss}{OT1}{cmss}{bx}{n}
  \SetMathAlphabet\mathbfss{bold}{OT1}{cmss}{bx}{n}
  \ifAMStwofonts
    \ifCUPmtlplainloaded \else
      \DeclareSymbolFont{UPM}{U}{eur}{m}{n}
      \SetSymbolFont{UPM}{bold}{U}{eur}{b}{n}
      \DeclareSymbolFont{AMSa}{U}{msa}{m}{n}
      \DeclareMathSymbol{\upi}{0}{UPM}{"19}
      \DeclareMathSymbol{\umu}{0}{UPM}{"16}
      \DeclareMathSymbol{\upartial}{0}{UPM}{"40}
      \DeclareMathSymbol{\leqslant}{3}{AMSa}{"36}
      \DeclareMathSymbol{\geqslant}{3}{AMSa}{"3E}

      \let\leq=\leqslant 

    \fi
  \fi
\fi 

\ifCUPmtlplainloaded \else
  \ifAMStwofonts \else 
    \def\upi{\pi}
    \def\umu{\mu}
    \def\upartial{\partial}
  \fi
\fi

\title{Predicted Planck Extragalactic 
Point Source Catalogue}
\author[Vielva et al.]
       {P.~Vielva$^{1,2,}$\footnotemark,
	E.~Mart{\'\i}nez-Gonz{\'a}lez$^{1}$,
        L.~Cay{\'o}n$^{1}$,
	J.~M.~Diego$^{1,2}$
	J.~L.~Sanz$^{1}$ and \and
        L.~~Toffolatti$^{3}$\\
	$^{1}$Instituto de F{\'\i}sica de Cantabria, Fac. Ciencias,
	Avda. Los Castros s/n, 39005, Spain \\
        $^{2}$Departamento de F{\'\i}sica Moderna, Universidad de Cantabria,
        Avda. Los Castros s/n, 39005 Santander, Spain\\
	$^{3}$Departamento de F{\'\i}sica, Universidad de Oviedo,
        c/ Calvo Sotelo s/n, 33007 Oviedo, Spain}
\date{\today}

\pagerange{\pageref{firstpage}--\pageref{lastpage}}
\pubyear{1994}

\begin{document}

\maketitle

\label{firstpage}

\begin{abstract}
An estimation of the number and amplitude (in flux) of the
extragalactic point sources that will be observed by the
Planck Mission is presented in this paper. The study is based
on the Mexican Hat wavelet formalism introduced by
Cay{\'o}n et al. 2000. Simulations at Planck observing frequencies
are analysed, taking into account all the possible cosmological,
Galactic and Extragalactic emissions together with noise.
With the technique used in this work the Planck Mission will
produce a catalogue of extragalactic point sources above fluxes:
1.03 Jy (857 GHz), 0.53 Jy (545 GHz), 0.28 Jy (353 GHz),
0.24 Jy (217 GHz), 0.32 Jy (143 GHz), 0.41 Jy (100 GHz HFI),
0.34 Jy (100 GHz LFI), 0.57 Jy (70 GHz), 0.54 Jy (44 GHz)
and 0.54 Jy (30 GHz), which are only slightly model dependent
(see text). Amplitudes of these sources are estimated with errors
below $\sim 15\%$. Moreover, we also provide a complete catalogue
(for the point sources simulation analysed) with errors in the
estimation of the amplitude below $\sim 10\%$. In addition we discuss
the possibility of identifying different point source populations
in the Planck catalogue by estimating their spectral indices.

\end{abstract}


\section{Introduction}
\footnotetext{e-mail: vielva@ifca.unican.es}
The most accurate observations of the Cosmic Microwave Background
(CMB) covering angular scales down to a few arcmins will be
provided by the future ESA Mission: Planck (Mandolesi et al. 1998,
Puget et al. 1998). Two instruments, the LFI
(Low Frequency Instrument) and the HFI (High Frequency Instrument)
will produce 10 maps of the sky at 9 frequencies ranging from
30 GHz to 857 GHz. High sensitivity together with high resolution will
characterize these observations. Several emissions will contribute
to the maps. The CMB signal and instrumental noise will be observed
together with Galactic (free-free, synchrotron and dust emissions)
and extragalactic foregrounds (extragalactic point sources and
Sunyaev-Zel'dovich sources) in the Planck channels. One major
concern is to develop methods to separate all these foregrounds
from the cosmological signal. Several methods have already been
proposed (Hobson et al. 1998, Bouchet et al. 1999,
Baccigalupi et al. 2000) and there is a lot of work left to be done.

In this paper we focus on the extraction of extragalactic point
sources from simulated Planck maps. Several methods
have been developed to extract point sources from CMB maps as
those presented in Hobson et al. 1999, Tegmark and
Oliveira-Costa 1998 (TOC98, hereafter), Tenorio et al. 1999 and
Cay{\'o}n et al. 2000 (C00, hereafter). Hobson et al. based their work
on Maximum Entropy Methods (MEM) to separate all the foregrounds
in simulated Planck maps. They obtained residual maps in which
instrumental noise plus extragalactic point sources were still
presents. Point sources were finally extracted from the maps using
SExtractor. TOC98 aimed to extract point sources from Planck data
using a filter designed to amplify point source emissions by
reducing the rms of other emissions. This method relies on the
Gaussianity of the CMB and foregrounds and requires some knowledge
about the power spectrum of each of the sources in a map. The last
two papers present methods based on wavelets to extract point
sources from microwave maps. More realistic simulations of the
signals and point sources that will be observed in future microwave
maps were used in C00. Moreover, the most important difference
between the two wavelet based works is the wavelet used in the
analysis. In a recent work, Sanz et al. 2000, show that the
Mexican Hat wavelet (MHW hereafter) is the optimal pseudofilter
to detect point sources in microwave maps, at least for a wide
range of theoretical power spectra. It is also important to
notice that the method described in C00 does not introduce any
assumptions about the nature of the CMB or other foregrounds.

This paper continues the method introduced in C00 to extract point
sources from microwave maps. We present a complete study of Planck
simulations (at all observing frequencies) to construct a Planck
Extragalactic Point Source Catalogue. The paper is organized as
follows. In Section 2 we describe the simulations to be used.
The MHW formalism is briefly reviewed in Section 3, and a discussion
about the optimal MHW scale to detect point sources in the different
Planck channels is also included. The detection and flux estimation 
results are presented in Section 4. Section 5 includes a discussion
on the possible estimation of spectral characteristics of different
point source populations. Finally, the conclusions of this work are
presented in Section 6.

\section{Planck Simulations at different frequencies}
In order to predict how many extragalactic point sources
will be observed in the Planck maps, we have simulated
those maps including all the experimental constraints
(given in Table 1) as well as all the emissions at each
frequency. Although Planck will produce whole sky maps,
we perform our analysis on squared regions of the sky
of $12.8^{\circ}\times 12.8^{\circ}$. This procedure of dividing the
celestial sphere in small, almost flat, patches does not 
imply any lost of generality since the scales relevant for
the source detection are in the arcmin range. We restrict
our analysis to regions outside the Galactic plane,
defined by $-30^{\circ} < b < +30^{\circ}$. This sky cut is quite
conservative and should be frequency dependent;
for simplicity we will use the same cut for all maps.
The total number of point sources to be observed by
Planck will be given by the addition of the ones detected
in squared regions covering a total area of
$2\pi$ sr (half of the sphere).

The Galactic emission in the total area of the sky we analyse
changes (depending on the observing frequency) from one
squared region to another. To take into account the possible
variations we have considered several representative squared
regions characterized by different Galactic emission intensities.
Specifically, we have studied ten different
$12.8^{\circ}\times 12.8^{\circ}$
sky patches, each of them randomly chosen from each of the
ten equally spaced intervals in which the dust emission
(outside the Galactic cut) can be tentatively divided.
We have chosen dust as the guide to divide the sky since it
is clearly the dominant emission at high Planck frequencies.
The other Galactic emissions are expected to be, in most of
the sky outside the Galactic cut, below either CMB or dust
emissions at all Planck frequencies. We have analysed the
same ten patches at all Planck frequencies, and according to
the results, we can distinguish four different zones in the
857 GHz channel, three in 545, 353 and 217 GHz channels, and
just two in 143, 100, 70, 44 and 30 GHz channels. The
characteristic values for each Galactic emission in these
zones are given in Table 2. Thereby, six different patches
can be assigned to Zone I at 857 GHz, two different ones at
Zone II, and only one at both Zone III and Zone IV. At Zone
I of the three next channels (545, 353 and 217) eight
different patches can be assigned and only one at
Zones II and III. Finally, at Zone I of the rest channels
we assign nine different patches and just one at Zone II.

\begin{table}
   \begin{center}
      \label{table:CaracTecni}
         \begin{tabular}{|c|c|c|c|}
	 \hline
	 Frequency & FWHM & Pixel size & $\sigma_{noise}$ \\
	 (GHz) & (arcmin) & (arcmin) & $(10^{-6})$ \\
	 \hline
	 857 & 5.0 & 1.5 & 22211.10 \\
	 \hline
	 545 & 5.0 & 1.5 & 489.51 \\
	 \hline
	 353 & 5.0 & 1.5 & 47.95 \\
	 \hline
	 217 & 5.5 & 1.5 & 15.78 \\
	 \hline
	 143 & 8.0 & 1.5 & 10.66 \\
	 \hline
	 100 (HFI) & 10.7 & 3.0 & 6.07 \\
	 \hline
	 100 (LFI) & 10.0 & 3.0 & 14.32 \\
	 \hline
	 70 & 14.0 & 3.0 & 16.81 \\
	 \hline
	 44 & 23.0 & 6.0 & 6.79 \\
	 \hline
	 30 & 33.0 & 6.0 & 8.80 \\
	 \hline
      \end{tabular}
      \caption{Experimental constrains at the 10 Planck channels. The
      antenna FWHM is given in column 2 for the different frequencies
      (a Gaussian pattern is assumed in the HFI and LFI
      channels). Characteristic pixel sizes are shown in column 3. The
      fourth column contains information about the instrumental noise
      level, in $\Delta T/T$ per pixel.}
    \end{center}
\end{table}

Dust emission have been simulated using the data and the model
provided by Finkbeiner et al. 1999. This model assumes that
dust emission is due to two components: a \emph{hot} one with a mean
dust temperature of ${T_D}^{hot} \simeq 16.2K$ and an
emissivity $\alpha^{hot} \simeq 2.70$, and a \emph{cold} one with a
${T_D}^{cold} \simeq 9.4K$, and an $\alpha^{cold} \simeq 1.67$.

Free-free emission is poorly known.
Present experiments such as H-$\alpha$ Sky Survey
\footnote{http://www.swarthmore.edu/Home/News/Astronomy/}
and the WHAM project \footnote{http://www.astro.wisc.edu/wham/}
will provide maps
of $H_{\alpha}$ emission that could be used as a template
for this emission.
As a first approximation we have simulated the
emission due to free-free using the correlation with
dust emission in the way proposed by Bouchet et al. 1996.

Synchrotron emission simulations have been done using the
all sky template provided by P. Fosalba and G. Giardino
in the ftp site: \emph{ftp://astro.estec.esa.nl}. This map
is an extrapolation of the 408 MHz radio map of Haslam et al.
1982, from the original $1^{\circ}$ resolution to a resolution
of about $5$ arcmin. 
A power law for the power spectrum 
with an exponent of $-3$ has been assumed. 
We have done an additional
extrapolation to the smallest scale ($1.5$ arcmin)
following the same power law.
We include in our simulations the information on the changes of 
spectral index as a function of electron density in the Galaxy.
This template have been done combining the Haslam
map with the Jonas et al. 1998 at 2326 MHz and with the Reich
\& Reich 1986 map at 1420 MHz, and can be found in the previous
ftp site.

\begin{table*}\label{table:OptimalScales}
   \begin{center}
      \begin{tabular}{|c|c|c|c|c|c|}
          \hline
          Frequency & Optimal Scale & Dust emission &
          CMB emission & Free-Free & Synchrotron emissions\\
	  (GHz) & ($R/\sigma_a$) & ($\Delta T/T$) & ($\Delta T/T$) &
	  ($\Delta T/T$) & ($\Delta T/T$) \\
 	  \hline
	  857, Zone I & 1.0 & $9.03\times 10^{-2}$ \dag & $4.38\times 10^{-5}$ &
                                $6.91\times 10^{-5}$ & $5.03\times 10^{-5}$ \\
	  857, Zone II & 0.9 & $1.40\times 10^{-1}$ \dag & " &
                                $3.96\times 10^{-5}$ & $1.73\times 10^{-5}$\\
	  857, Zone III & 0.8 & $5.48\times 10^{-1}$ & " &
                                $9.27\times 10^{-5}$ & $8.30\times 10^{-5}$\\
	  857, Zone IV & 0.6  & $1.55$ & " &
                                $4.91\times 10^{-4}$ & $2.37\times 10^{-5}$\\
	  \hline
	  545, Zone I & 1.3  & $6.94\times 10^{-4}$ \dag & $4.38\times 10^{-5}$ &
                                $1.92\times 10^{-6}$ & $1.56\times 10^{-6}$ \\
	  545, Zone II & 0.9 & $4.54\times 10^{-3}$ & " &
                                $2.58\times 10^{-6}$ & $2.46\times 10^{-6}$\\
	  545, Zone III & 0.6 & $1.22\times 10^{-2}$ & " &
                                $1.37\times 10^{-5}$ & $7.63\times 10^{-7}$\\
	  \hline
	  353, Zone I & 1.2 & $3.67\times 10^{-5}$ \dag & $4.38\times 10^{-5}$ &
                                $4.04\times 10^{-7}$ & $3.63\times 10^{-7}$\\
	  353, Zone II & 1.0  & $2.54\times 10^{-4}$ & " &
                                $5.43\times 10^{-7}$ & $5.49\times 10^{-7}$\\
 	  353, Zone III & 0.7 & $6.54\times 10^{-4}$ & " &
                                $2.87\times 10^{-6}$ & $1.85\times 10^{-7}$\\
	  \hline
	  217, Zone I & 0.9 & $3.25\times 10^{-6}$ & $4.35\times 10^{-5}$ &
                                $2.71\times 10^{-7}$ & $2.73\times 10^{-7}$\\
	  217, Zone II & 0.9 & $3.38\times 10^{-5}$ & " &
                                $3.64\times 10^{-7}$ & $3.94\times 10^{-7}$\\
	  217, Zone III & 0.8 & $8.41\times 10^{-5}$ & " &
                                $1.92\times 10^{-6}$ & $1.46\times 10^{-7}$\\
	  \hline
	  143, Zone I & 0.7 & $1.01\times 10^{-6}$ & $4.21\times 10^{-5}$ &
                                $3.70\times 10^{-7}$ & $4.11\times 10^{-7}$\\
	  143, Zone II & 0.7 & $2.55\times 10^{-5}$ & " &
                                $2.57\times 10^{-7}$ & $2.28\times 10^{-7}$\\
	  \hline
	  100 (HFI), Zone I & 0.6 & $4.60\times 10^{-7}$ & $4.05\times 10^{-5}$ &
                                $6.23\times 10^{-7}$ & $7.49\times 10^{-7}$\\
	  100 (HFI), Zone II & 0.6 & $1.14\times 10^{-5}$ & " &
                                $4.23\times 10^{-6}$ & $4.31\times 10^{-7}$\\
	  \hline
	  100 (LFI), Zone I & 0.8 & $4.59\times 10^{-7}$ & $4.09\times 10^{-5}$ &
                                $6.24\times 10^{-7}$ & $7.51\times 10^{-7}$\\
	  100 (LFI), Zone II & 0.8 & $1.13\times 10^{-5}$ & " &
                                $4.21\times 10^{-6}$ & $4.32\times 10^{-7}$\\
	  \hline
	  70, Zone I & 0.8 & $2.32\times 10^{-7}$ & $3.87\times 10^{-5}$ &
                                $1.18\times 10^{-6}$ & $1.54\times 10^{-6}$\\
	  70, Zone II & 0.7 & $5.57\times 10^{-6}$ & " &
                                $7.77\times 10^{-6}$ & $9.17\times 10^{-7}$ \\
	  \hline
	  44, Zone I & 0.7 & $1.02\times 10^{-7}$ & $3.43\times 10^{-5}$ &
                                $2.97\times 10^{-6}$ & $4.17\times 10^{-6}$\\
	  44, Zone II & 0.7  & $2.28\times 10^{-6}$ & " &
                                $1.82\times 10^{-5}$ & $2.60\times 10^{-6}$\\
	  \hline
	  30, Zone I & 0.6 & $5.26\times 10^{-8}$ & $3.03\times 10^{-5}$ &
                                $6.53\times 10^{-6}$ & $9.74\times 10^{-6}$\\
	  30, Zone II & 0.6 & $1.13\times 10^{-6}$ & " &
                                $3.84\times 10^{-5}$ & $6.25\times 10^{-6}$\\
	  \hline
      \end{tabular}
      \caption{MHW optimal scales for the Planck channels in 
       different Zones of the sky. The rms values
       of the Galactic and CMB emissions are also presented
       at each of the selected zones. \dag The dust amplitude
       intervals for these zones, in $\Delta T/T$, are:
       $5.54\times 10^{-2}$ -- $1.35\times 10^{-1}$,
       857 GHz Zone I;
       $1.40\times 10^{-1}$ -- $2.46\times 10^{-1}$,
       857 GHz Zone II;
       $4.33\times 10^{-4}$ -- $1.03\times 10^{-3}$,
       545 GHz, Zone I;
       $2.31\times 10^{-5}$ -- $5.46\times 10^{-5}$,
       353 GHz, Zone I.}
   \end{center}
\end{table*}

We have also taken into account the possible Galactic emission
due to rotational grains of dust, proposed by Draine
\& Lazarian 1998. This component could be
important at the lowest frequencies of the Planck channels
(30 and 44 GHz) in the outskirts of the Galactic plane, where it is
around two times lower than the free-free emission.
As the authors propose in that paper, this emission is correlated
with the thermal dust one, through the neutral hydrogen
column density ($N_H$):

\begin{equation}
{I(\nu)}_{rot} = f(\nu)N_H, \hspace{0.5cm}
{I(3000~GHz)_{thermal}} = a N_H,
\nonumber
\end{equation}
where $f(\nu)$ is the frequency dependence of the emissivity
predicted by Draine \& Lazarian
and $a$ is the correlation between the $21~cm$ emission and the
infrared dust one. We addopt the correlation proposed by Boulanger
\& P{\'e}rault (1988):
\begin{equation}
a \approx 0.85\times 10^{-14} Jy~ sr^{-1}
{\Big(\frac{H~atoms}{cm^{-2}}\Big)}^{-1}.
\nonumber
\end{equation}
Hence, the rotational dust emission is simulated from the thermal
one through the equation:
\begin{equation}
{I(\nu)}_{rot} =  a^{-1} f(\nu) {I(3000~GHz)_{thermal}}.
\nonumber
\end{equation}

Sunyaev-Zel'dovich (SZ) effect simulations developed by
Diego et al. 2000 have also been added to the maps. 
These simulations assume a flat $\Lambda$CMD Universe with
$\Omega_m = 0.3$ and $\Omega_{\Lambda}$ = $0.7$.
The CMB emission have been simulated for the same Universe,
using the \emph{Cl's} generated with the CMBFAST
code (Seljak \& Zaldarriaga, 1996).

Finally, the extragalactic point source simulations have been 
performed following the model of Toffolatti et al. 1998
(see their paper for more details) assuming the cosmological model
indicated above.

\section{Mexican Hat Wavelet method: detection and amplitude
estimation of point sources}
Only recently have wavelet techniques been applied to 
analyze CMB maps.
Among the different applications, they have been used to
denoise CMB maps
(Sanz et al. 1999a, 1999b, Tenorio et al. 1999), to detect
non-Gaussianity (Ferreira et al. 1997, Hobson et al. 1999,
Aghanim and Forni 1999) and, more recently, to detect and subtract
point sources (C00, Tenorio et al. 1999).

\begin{figure*}\label{fig:OptSca}
   \begin{center}
   \epsfysize=160mm
   \epsfxsize=170mm
   \epsffile{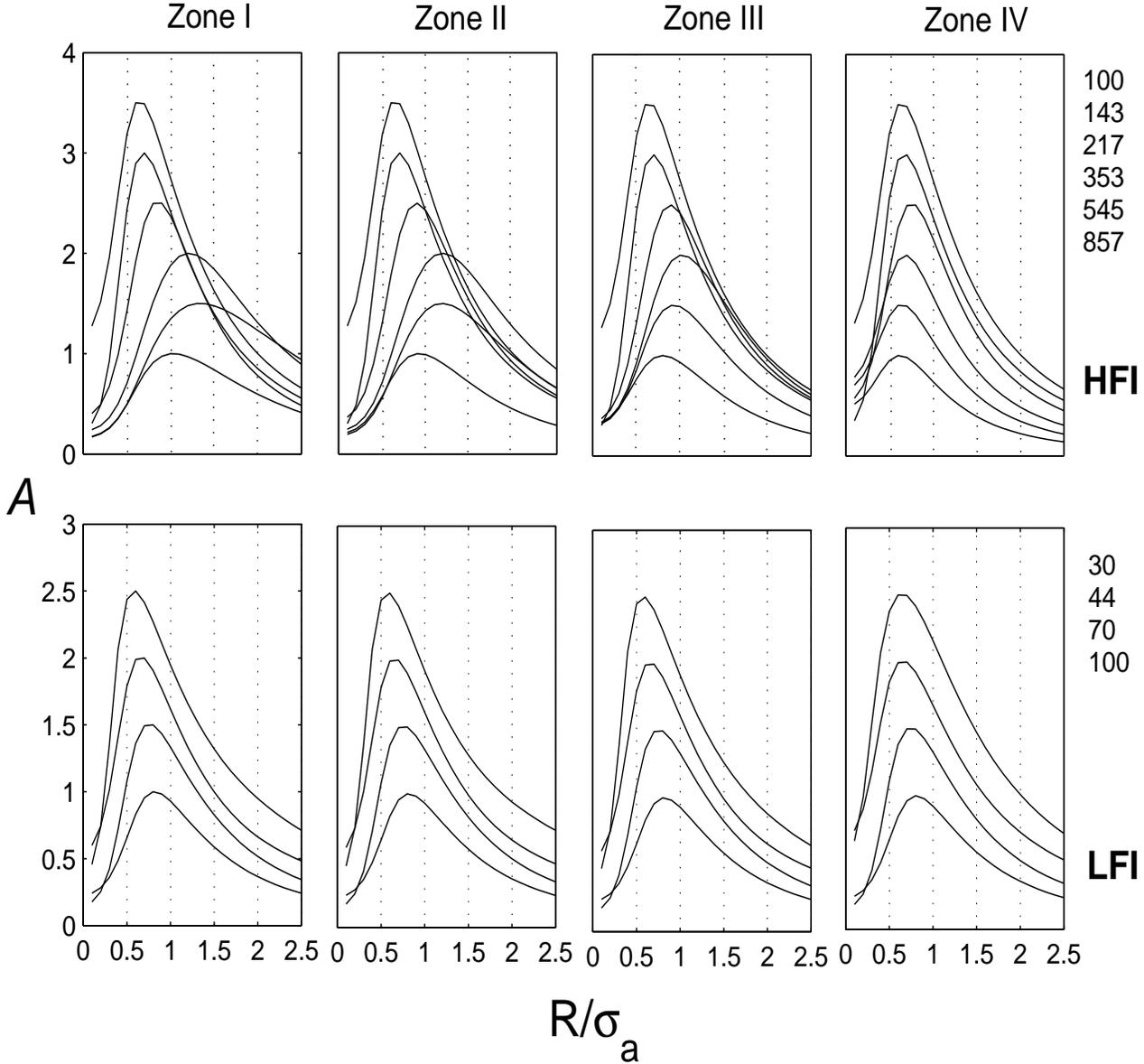}
   \caption{Amplification 
    (arbitrary units) as a function of Mexican Hat scale
    in units of the antenna dispersion at each frequency,
    for different selected zones of the sky.}
   \end{center}
\end{figure*}

\begin{figure*}\label{fig:Mapas}
\begin{center}
\epsfxsize=140mm
\epsffile{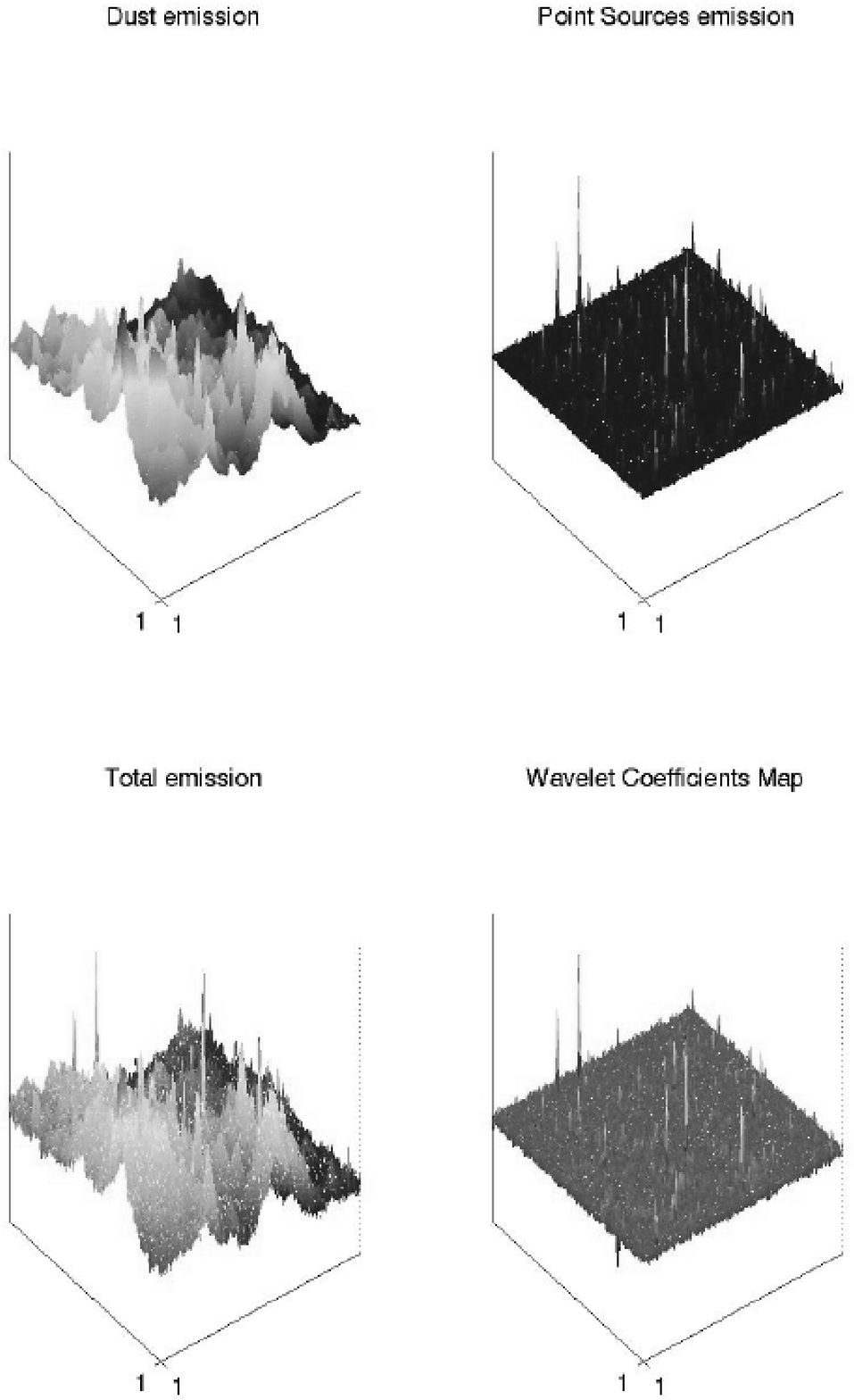}
\caption{A graphical example of how the MHW works
for one of our simulations at 857 GHz.
The top left figure shows the dust emission.
Point Sources are presented in the top right figure.
The total emission, with all
the Galactic and Extragalactic emissions, as well as the cosmological
signal and the instrumental noise, is shown in the bottom left
figure. The bottom right figure shows
the wavelet coefficient map at the optimal scale.}
\end{center}
\end{figure*}

We will study the detection of 
point sources in Planck simulated maps using the method
introduced by C00. This method is based on the adequacy of the 
Mexican Hat to select point sources with a characteristic 
Gaussian shape introduced by the antenna beam.
The MHW is an isotropic wavelet,
characterized by a scale $R$. The analytical expression is given by
\begin{equation}\label{eq:MHW}
   \psi(x) = \frac{1}{\sqrt{2\pi}}\Big[2-\big(\frac{x}{R}\big)^2\Big]
             e^{-\frac{x^2}{2R^2}}.
\end{equation}
Wavelet analyses are based on the study of the so called
wavelet coefficients obtained by a convolution of the 
signal (a 2-D map in this case) being analyzed, with the wavelet. 
Thus, wavelet coefficient maps can be obtained at each characteristic
scale $R$. In the case of a point source at position $\vec x_o$ of
amplitude $B$ and convolved with a Gaussian of dispersion $\sigma_a$, one 
can easily calculate the analytical expression of the
wavelet coefficient:  

\begin{equation}\label{eq:Coeff}
   \frac{w(R)}{R} = 2\sqrt{2\pi}\frac{B}{A}
                              \frac{(R/\sigma_a)^2}
                              {(1 + (R/\sigma_a)^2)^2},
\end{equation}
where $A$ is the area occupied by the point source (this is 
introduced for normalization purposes).

We are interested in detecting point sources from a background
of CMB, noise and foreground emissions. The 
MHW increases the amplitude of the
wavelet coefficients relative to the dispersion of the
wavelet coefficient map $\sigma_w(R_{opt})$, 
at point source positions, at a certain optimal scale $R_{opt}$:
\begin{equation}\label{eq:Ampl}
   \frac{w(R_{opt})}{\sigma_{w}(R_{opt})} > \frac{(B/A)}{\sigma},
\end{equation} 
where $\sigma$ is the dispersion of the observed map (dispersion
in real space).
The dispersion of the wavelet coefficients can be calculated from:
\begin{equation}\label{eq:SW}
   \sigma_{w}(R) = 2\pi R^{2}\int dkP(k)k |\widehat{\psi}(Rk)|^2,
\end{equation}
where $P(k)$ correspond to the power spectrum of the map
to be analyzed and $\widehat{\psi}$ is the Fourier transform
of the MHW.
This together with the analytical expression of the
wavelet coefficient allow us to estimate the 
amplification defined by:

\begin{equation}\label{eq:Amplifi}
   \textrm{$\mathcal{A}$} \equiv \frac{w(R_{opt}) \sigma}
                             {(B/A)\sigma_{w}(R_{opt})}.
\end{equation}

The amplification reaches its maximum value at the optimal
scale. As it will be shown below, the optimal
scale is defined by a combination of 
the characteristic scale and intensity of the foreground
(or CMB background) that dominates 
the map and the amount of instrumental noise. We have calculated the optimal
scale for each of the characteristic zones defined in the previous 
section at each frequency. The results are 
presented in Figure 1 and Table 2. Values of the
optimal scale for each zone are given in that table. The variation
of the optimal scale inside each zone is less than 5\%.

As expected the optimal scale takes values around the antenna
dispersion becuase this scale is the one that characterizes a point source.
At low frequency the main emission is due to CMB. The
typical CMB \emph{coherence scale} is $\simeq 10'$, equal
or smaller than  the antenna dispersion at those frequencies.
In this situation both point sources and background have similar
variation scales (although the background characteristic scale
will always be greater than the point source one). The optimal
scale tends to be smaller than the antenna dispersion to better 
liminate the background contribution. Zones I and II at high
frequencies will need optimal scales larger than $\sigma_a$ to extract
the point sources. This is due to the importance of noise relative
to the dust emission. At zones strongly dominated by dust
(Zones III and IV), although the background has a characteristic
scale much larger than the antenna dispersion, however the optimal
scale must be below $\sigma_a$. This is necessary to better erase the dust
features (notice that in these zones the noise amplitude is much 
lower than the dust one, representing a minor problem). Therefore,
the optimal scale on each map is a compromise between the background
(its variation scale and its intensity) and the noise, in the
sense that it tends to compensate for the effect of the most
harmful contribution due to either the background or the noise.

We have plotted in Figure 2 an example of the effect of applying
the MHW to a dust dominated map. At the optimal scale one can
clearly see the point sources selected. Comparison of the two
panels on the right gives a very good idea of how powerful the
MHW is to detect point sources in those maps.

Once the optimal scale is calculated, we perform the point source 
detection on the corresponding wavelet map. We select $k$ positions
with wavelet coefficient $w(R_{opt})$ values
larger than a certain threshold.
Moreover, to make sure that the detection
is a point source we fit the ``experimental'' $w(R,\vec x_o)$
versus $R$ curve (obtained from the simulations)
to the expected theoretical value given by eq. (5). In this paper 
we improve the $\chi^2$ method used in C00. Correlations between the
points in that curve were ignored in C00. To take this into account we define
the $\chi^2$ at pixel $k$ by
\begin{equation}\label{eq:Chi2}
   {\chi^2}_k = \sum_{i,j} ({w_{i,k}}^{theo} -
   {w_{i,k}}^{exp}){V_{i,j}}^{-1}
   ({w_{j,k}}^{theo} - {w_{j,k}}^{exp}),
\end{equation}
where $V_{i,j}$ are elements of the covariance
matrix between the different scales $(i,j)$ and ${w_{i,k}}^{exp}$
and ${w_{i,k}}^{theo}$ represent the experimental and theoretical
values of the wavelet coefficients of the selected pixels $(k)$
at scale $i$.
The covariance between scales $i$ and $j$ is given by
\begin{equation}\label{eq:CorrelationMatrix}
V_{ij} =\frac{1}{N} \sum_{k = 1}^{N}
{w_{i,k}}^{exp}{w_{j,k}}^{exp},
\end{equation}
where $N$ is the number of wavelet coefficients at each scale
(the number of pixels in the map). On our analysis we use
a four scales fit, so $\mathbf{V}$ is a $4\times4$ matrix.
This is an approximated covariance matrix,
since it should be formed by the \emph{noise}
wavelet coefficients. Here, the \emph{noise} must be understood
as all the components except the point source which is being
fitted. But our aim is, preciselly, to estimate this
point source. Hence, we are not able to determine,
a priori, the point source contribution to
that coefficient. However, the covariance matrix deffined
by eq. (10) is approximately the exact one,
since the contribution of each individual point source wavelet
coefficient to the covariance matrix is negligible.

Two different criterions will be applied. In the first one,
only pixels above $5\sigma_w(R_{opt})$ in the wavelet map at
$R_{opt}$ and with acceptable reduced $\chi^2$ fits will
be the ones selected. Taking into account our simulations,
a reduced $\chi^2$ is acceptable if it is lower or equal than 4.
However, we accept a detection with a reduced $\chi^2$ greater
than the previous value if this detection appears in an adjacent
channel with an acceptable reduced $\chi^2$.
This allows us to include in our catalogue some detections with
low errors but without a total satisfactory fit. The amplitude of
the point sources detected in this way is calculated from this fit.
The second criterion is based on the error in the amplitude
estimation. We look for those flux limits such that the errors
are $\leq 50\%$ with a maximum percentage of spurious detections
(i.e. error $> 50\%$) lower than 5\%. We consider only wavelet
coefficients above $2\sigma_w(R_{opt})$. As in the first criterion,
we determine the amplitude from the fitted profile.
We use a maximum $\chi^2$ limit as in the previous criterion,
with the same exception. We have applied these criterions
to Planck simulations (described in Section 2) to generate
the point source catalogues described in the following section.

\section{Point Source Catalogue}
Taking into account the number of different squared
regions that are in the area of the sky analysed
(outside the Galactic plane), we present in this
Section the results of applying the MHW method (and
the two different criterions) to obtain point source
catalogues from the simulated Planck maps. We present
two Tables for each criterion: one with the lowest
fluxes achieved and other with the 100\% completeness fluxes.
\begin{table*}
\begin{minipage}{170mm}
\begin{center}
\textbf{Point Source detections in $2\pi$ sr. $5\sigma_w(R_{opt})$ criterion.}
\end{center}
\begin{center}
\begin{tabular}{|c|c|c|c|c|c|c|c|c|}
\hline
Frequency (GHz) &
$N(>5\sigma_{wc})$ &
\% Bad detections &
$\overline{E}_{abs}(\%)$ &
$\overline{E}(\%)$ &
Ampl. &
Min flux (Jy) &
\% Detections &
$N(>5\sigma_{rs})$ \\
\hline
857 & 9408 & 1.5 & 9.9 & -4.6 & 31.50 & 1.03 & 38.0 & 1907 (734) \\
\hline
545 & 2427 & 3.7 & 13.8 & -8.9 & 17.44 & 0.53 & 33.5 & 1900 (445) \\
\hline
353 & 1088 & 2.6 & 14.3 & -7.2 & 5.20 & 0.28 & 72.7 & 654 (250) \\
\hline
217 & 605 & 0.2 & 9.0 & -4.1 & 6.85 & 0.24 & 74.5 & 26 (0) \\
\hline
143 & 651 & 0.1 & 9.5 & -0.9 & 5.01 & 0.32 & 53.9 & 23 (16) \\
\hline
100 (HFI) & 600 & 0.3 & 8.2 & -4.4 & 3.62 & 0.41 & 58.9 & 6 (0) \\
\hline
100 (LFI) & 486 & 0.1 & 11.3 & -4.7 & 3.31 & 0.34 & 51.8 & 44 (39) \\
\hline
70 & 358 & 0.3 & 9.3 & -2.8 & 2.84 & 0.57 & 87.3 & 5 (0) \\
\hline
44 & 350 & 0.5 & 10.9 & -5.8 & 2.80 & 0.54 & 50.2 & 13 (5) \\
\hline
30 & 364 & 0.6 & 14.4 & -9.6 & 2.95 & 0.54 & 51.7 & 18 (6) \\
\hline
\end{tabular}
\end{center}
\caption{The frequency of each Planck channel is indicated in column
1. The second column shows the estimation of the number
of detections above $5\sigma_w(R_{opt})$. The percentage of
spurious detections with an error in the amplitude
estimation $> 100\%$ (those spurious detections are generally
contaminated by noise) is given in the third column. The absolute
and real values of the mean error associated to the amplitude
estimation are shown in columns 4 and 5. The mean amplification as
defined in (8) is given in column 6. The seventh column shows
the minimum flux reached. The percentage of point sources detected
out of the total number of point sources present in the simulation
is given in column 7. For comparison, the number of detections above
$5\sigma_{rs}$ in the observed maps (real space) are shown in
column 9. Numbers in brackets indicate the number of real point
sources out of the total detected ($\sigma_{rs}$ $(\Delta T/T)$ =
$1.13$ (857 GHz),
$8.51\times 10^{-4}$ (545 GHz),
$6.28\times 10^{-5}$ (353 GHz),
$4.37\times 10^{-5}$ (217 GHz),
$4.21\times 10^{-5}$ (143 GHz),
$4.05\times 10^{-5}$ (100 GHz, HFI),
$4.09\times 10^{-5}$ (100 GHz, LFI),
$3.87\times 10^{-5}$ (70 GHz),
$3.46\times 10^{-5}$ (44 GHz),
$3.18\times 10^{-5}$ (30 GHz)).}
\end{minipage}
\end{table*}

\subsection{First criterion: $5\sigma_w(R_{opt})$}
In Table 3, we give the results achieved with this criterions,
for the lowest fluxes we are able to reach. The total number
of point sources detected at each of the 10 Planck channels
is given in the second column. We just detect the point sources
in the tail of the flux number counts. We assume poissonian
error bars for the number of point source detections.
This is justified since for a given point source realization
the difference in the number of detections from one
$12.8^{\circ}\times 12.8^{\circ}$
patch of the sky to another in the same zone is
much smaller than the poissonian error. The percentage
of spurious detections is indicated in the third column.
Though spurious detection are included in the number of
detected sources, they are not in columns refering to the
error and flux columns. These few "bad detections" are
low amplitude point sources located at pixels with large
noise fluctuations. The point source is correctly located
by the MHW method but its estimated amplitude has a large
error, $> 100\%$ . As the table shows, this tends to happen
less often at low frequencies where only the higher amplitude
point sources are detected. In any case, the percentage of
these "bad detections" is small at all frequencies $(< 4\%)$.
The fourth column in Table 3 shows that the mean absolute error
in the amplitude estimation is always $< 15\%$. The fifth
column shows that there is a slight bias in the amplitude
estimation that tends to overestimate it. This method is
based on the amplification of the ratio
$w(R_{opt})/\sigma_w(R_{opt})$ relative to that ratio in
real space. As can be noticed from the sixth column in Table 3,
the amplification is usually larger at high frequencies where
dust is the dominating foreground. All the detections presented
in Table 3 are above the flux levels indicated in column 7.
Out of all the point sources existing in the maps above those
fluxes, the percentage detected is indicated in column 8.
For comparison we give in the last column of Table 3 the
number of point sources that will be detected in the original
simulated maps by selecting all the pixels above
$5\sigma$, where $\sigma$ is the \emph{rms} value of a
$12.8^{\circ}\times 12.8^{\circ}$ patch. Out of these pixels
selected, only the ones given within parenthesis are real
point sources. One has to take into account that there could
be maps with large gradients that will have large $\sigma$
values and will therefore not allow us to detect point
sources above $5\sigma$, even when they are visible.

It is also important to know the flux limit above which the
\emph{Planck Extragalactic Point Source Catalogue} will be
complete. This information is provided in Table 4. The
complete $(100 \%)$ \emph{Extragalactic Point Source Catalogue}
will contain a maximum of $\simeq 5000$ point sources in the 857
GHz channel, and a minimum of 300 point sources in the 30 GHz
channel. The error in the amplitude estimation is lower
than 10\% at all frequencies (except for the Zone II at 30
and 44 GHz, which represent $< 1\%$ of the sky outside the Galactic Plane).

\begin{table*}
\begin{minipage}{170mm}
\begin{center}
\textbf{Complete $(100 \%)$ Extragalactic Point Source Catalogue
in $2\pi$ sr. $5\sigma_w(R_{opt})$ criterion.}
\end{center}
\begin{center}
\begin{tabular}{|c|c|c|c|c|c|c|}
\hline
Frequency (GHz) &
$N(> Min Flux)$ &
$\overline{E}_{abs}(\%)$ &
$\overline{E}(\%)$ &
Amplification &
Min flux (Jy) &
\% Sky area \\
\hline
857, Zone I & 4531 & 6.0 & 0.2 & 18.79 & 1.73 & 91.8 \\
857, Zone II & 296 & 5.9 & 0.4 & 42.09 & 1.83 & 6.2 \\
857, Zone III & 37 & 6.5 & -0.3 & 56.95 & 2.18 & 1.0 \\
857, Zone IV & 0 & -- & -- & -- & -- & 1.0 \\
857, Whole Sky & 4864 & 6.0 & 0.2 & 20.63 & 1.74 & \\ 
\hline
545, Zone I & 1212 & 6.4 & -0.5 & 6.25 & 0.99 & 98.0 \\
545, Zone II & 12 & 6.6 & -2.2 & 44.63 & 1.13 & 1.0 \\
545, Zone III & 0 & -- & -- & -- & -- & 1.0 \\
545, Whole Sky & 1224 & 6.4 & -0.5 & 6.64 & 0.99 &\\
\hline
353, Zone I & 646 & 9.2 & 0.3 & 4.08 & 0.41 & 98.0 \\
353, Zone II & 6 & 9.5 & -0.4 & 61.6 & 0.42 & 1.0 \\
353, Zone III & 0 & -- & -- & -- & -- & 1.0 \\
353, Whole Sky & 652 & 9.2 & 0.3 & 4.66 & 0.41 &\\
\hline
217, Zone I & 466 & 7.2 & -1.1 & 14.34 & 0.32 & 98.0 \\
217, Zone II & 5 & 7.2 & -2.1 & 4.11 & 0.34 & 1.0 \\
217, Zone III & 2 & 5.6 & -0.1 & 6.03 & 0.51 & 1.0 \\
217, Whole Sky & 473 & 7.2 & -1.1 & 14.15 & 0.32 &\\
\hline
143, Zone I & 356 & 7.5 & 1.5 & 4.63 & 0.56 & 99.0 \\
143, Zone II & 4 & 6.9 & 1.4 & 5.76 & 0.61 & 1.0 \\
143, Whole Sky & 360 & 7.5 & 1.5 & 4.64 & 0.56 &\\
\hline
100 (HFI), Zone I & 417 & 6.0 & -1.9 & 3.55 & 0.63 & 99.0 \\
100 (HFI), Zone II & 4 & 5.8 & 0.4 & 3.88 & 0.61 & 1.0 \\
100 (HFI), Whole Sky & 421 & 6.0 & -1.9 & 3.55 & 0.63 &\\
\hline
100 (LFI), Zone I & 409 & 8.4 & -2.2 & 3.35 & 0.64 & 99.0 \\
100 (LFI), Zone II & 5 & 7.2 & -1.9 & 2.82 & 0.66 & 1.0 \\
100 (LFI), Whole Sky & 414 & 8.4 & -2.2 & 3.34 & 0.64 &\\
\hline
70, Zone I & 327 & 9.1 & -2.8 & 2.88 & 0.72 & 99.0 \\
70, Zone II & 4 & 9.4 & -1.4 & 2.62 & 0.77 & 1.0 \\
70, Whole Sky & 331 & 9.1 & -2.8 & 2.88 & 0.72 &\\
\hline
44, Zone I & 293 & 8.1 & -2.4 & 2.87 & 0.81 & 99.0 \\
44, Zone II & 2 & 11.8 & -1.4 & 2.62 & 1.02 & 1.0 \\
44, Whole Sky & 295 & 8.1 & -2.4 & 2.87 & 0.81 &\\
\hline
30, Zone I & 295 & 9.5 & -4.1 & 2.97 & 0.89 & 99.0 \\
30, Zone II & 2 & 26.6 & -26.6 & 2.50 & 0.89 & 1.0 \\
30, Whole Sky & 297 & 9.7 & -4.3 & 2.97 & 0.89 &\\
\hline
\end{tabular}
\caption{The number and characteristics of point sources
in the complete $(100 \%)$ catalogue are given in this table.
Results are separated for the different sky zones. We also give
an all sky estimation for the Complete Extragalactic Catalogue,
taking into account the amount of sky that each Zone represent.
Galactic emission increases from Zones I to IV (see Table 2).
The number of point sources in the complete catalogue is indicated
in the second column, these values have poissonian errors.
Absolute and real values of the amplitude estimation mean errors
are provided in columns 3 and 4. The mean amplification, as
defined in (8), is given in column 5. Column 6 shows the minimum
flux above which all point sources are detected.
The last column gives an estimation of the percentage of
the sky covered by each zone.}
\end{center}
\end{minipage}
\end{table*}

\begin{table*}
\centering
\begin{minipage}{170mm}
\begin{center}
\textbf{Point Source detections in $2\pi$ sr. $50\%$ error criterion.}
\end{center}
\begin{center}
\begin{tabular}{|c|c|c|c|c|c|c|c|c|}
\hline
Frequency (GHz) &
$N(>5\sigma_{wc})$ &
\% Bad detections &
$\overline{E}_{abs}(\%)$ &
$\overline{E}(\%)$ &
Min flux (Jy) &
\% Detections \\
\hline
857 & 24221 & 4.8 & 16.5 & -9.4 & 1.05 & 85.2 \\
\hline
545 & 3496 & 3.6 & 12.7 & -4.6 & 0.63 & 74.5 \\
\hline
353 & 1497 & 0 & 13.8 & -3.4 & 0.26 & 95.0 \\
\hline
217 & 1130 & 0 & 12.6 & -0.5 & 0.17 & 95.0 \\
\hline
143 & 1258 & 0 & 12.6 & -2.9 & 0.18 & 90.9 \\
\hline
100 (HFI) & 1526 & 0 & 13.5 & -5.9 & 0.19 & 80.8 \\
\hline
100 (LFI) & 1165 & 0 & 13.4 & -4.6 & 0.26 & 77.1 \\ 
\hline
70 & 1008 & 0 & 13.5 & -8.2 & 0.30 & 88.9 \\
\hline
44 & 990 & 0 & 15.3 & -9.8 & 0.40 & 95.0 \\
\hline
30 & 864 & 0 & 13.8 & -8.8 & 0.49 & 100.0 \\
\hline
\end{tabular}
\end{center}
\caption{The frequency of each Planck channel is indicated in column
1. The second column shows the estimation of the number of
detections with this criterion. The percentage of spurious detections
with an error in the amplitude estimation $> 100\%$ (those spurious
detections are generally contaminated by noise) is given in the
third column. The absolute and real values of the mean error 
associated to the amplitude estimation are shown in columns 4 and 5.
The minimum flux reached is shown in column 6. The percentage
of point sources detected out of the total number of point sources
present in the simulation is given in column 7.}
\end{minipage}
\end{table*}

\subsection{Second criterion: $50\%$ error}
In Table 5, we present the catalogue of point source obtained
above the lowest flux reached at each frequency. In the first
column, we give the number of point sources detected at each
channel. Results are extended to half of the sky. As explained
above, we assume these number to have poissonian errors.
The number of point sources detected are 2 to 3 times greater
than the ones founded by the other criterion (since
the $5\sigma_w(R_{opt})$ is quite conservative). In the second
column, we give the number of spurious detections, fixed to be
lower than $5\%$ by the criterion. In this case, the percentage
of "bad detections" is included in all the columns,
since the errors are not so large than the ones in the
other criterion. As in Table 3, the mean amplitude estimation
errors are given in columns 4 and 5. The absolute error in
the amplitude estimation is always lower than $16\%$. Actually,
the mean error provided by this criterion is greater than the one
from the first criterion, since we are detecting more sources with
lower fluxes. The first criterion provide, in this sense, a
\emph{robust} catalogue. The flux limits and the percentage of
simulated point sources above those fluxes is given in columns 6
and 7. The minimum fluxes achieved are, generally, lower than those
obtained with the $5\sigma_w(R_{opt})$ criterion. The best improvement
in the flux limit appears in the intermediate and low frequency
channels. In these cases the number of point sources detected
by the first criterion in each simulation (small patch) was
very small $(\leq 6)$. This means that we detect point sources
in the tail of the flux distributions. The new criterion increases
the number of detections which implies significant variations in the
flux limits achieved. This is not the case for the 857,
545\footnote{In this channel, the minimum flux limit achieved in
the first detection method is lower than in the second one.
This is due to faint sources highly amplified and appears
above $5\sigma_w$. These sources are too faint to be detected
above the flux threshold determined by the second method}
and 353 GHz channels. The number of point sources
detected with both methods gives a good determination of the
detection level (an increment of detections does not change 
significantly the flux limit).

Finally, in Table 6 we present the complete $(100 \%)$
\emph{Extragalactic Point Source Catalogue} obtained with this
criterion. It will contain a maximum of $\simeq 14000$ point
sources in the 857 GHz channel and a minimum of $\simeq 800$
sources in the lowest frequency channels. The error in the
amplitude estimation is lower than $15\%$ except for the 44 GHz
channel and Zone II at 30 GHz. Again, these errors are greater 
han the ones provided by the first criterion, whereas the number
of point sources detected is greater and the completeness fluxes
are lower. Another advantage of this method is that the division in
Zones of the different parts of the sky are less important. That
division into different Zones was due to the different amount of
amplification of the point sources. However the second criterion
does not depend so much on the amplification (the detection
procedure requires only to be above $2\sigma_w(R_{opt})$ threshold).

Finally, the results obtained from both criterions are clearly model
dependent, since we adopt a specific evolution model. However, if
we change the evolution model for dusty galaxies, by adopting
realistic models in agreement with current determinations of
the diffuse far-IR/sub-mm back ground (FIRB), we still find
source detection limits which are very similar to the quoted ones.
For example, by using model E of Guiderdoni et al. (1998)
the detection limit at 353 GHz does not change appreciably.
Confusion noise from undetected sources is increased by a factor
of $\sim 3$ but the dominant sources of noise at 353 GHz are
still dust emission from our Galaxy and the CMB itself.
On the other hand, the number of detected sources is obviously
higher. At higher frequencies galactic dust emission dominates
all the other noise sources even at high galactic latitudes
(adopting the template of Finkbeiner et al., 1999; see Table 2),
whereas at $\nu \leq 200$ GHz confusion noise from undetected
sources is always below either CMB and galactic noise."

\begin{table*}
\begin{minipage}{170mm}
\begin{center}
\textbf{Complete $(100 \%)$ Extragalactic Point Source Catalogue
in $2\pi$ sr. $50\%$ error criterion.}
\end{center}
\begin{center}
\begin{tabular}{|c|c|c|c|c|c|}
\hline
Frequency (GHz) &
$N(> Min Flux)$ &
$\overline{E}_{abs}(\%)$ &
$\overline{E}(\%)$ &
Min flux (Jy) &
\% Sky area \\
\hline
857, Zone I & 13089 & 11.1 & -2.5 & 1.19 & 91.8 \\
857, Zone II & 781 & 10.8 & -1.8 & 1.23 & 6.2 \\
857, Zone III & 77 & 10.3 & -1.5 & 1.52 & 1.0 \\
857, Zone IV & 3 & 2.4 & 1.5 & 6.90 & 1.0 \\
857, Whole Sky & 13950 & 11.0 & -2.4 & 1.25 \\
\hline
545, Zone I & 2803 & 11.2 & -2.5 & 0.67 & 98.0 \\
545, Zone II & 21 & 10.9 & -3.5 & 0.73 & 1.0  \\
545, Zone III & 1 & 2.6 & 0.4 & 2.58 & 1.0 \\
545, Whole Sky & 2825 & 11.1 & -2.4 & 0.69 \\
\hline
353, Zone I & 1235 & 13.9 & -1.9 & 0.28 & 98.0 \\
353, Zone II & 10 & 13.7 & -1.8 & 0.30 & 1.0 \\
353, Zone III & 1 & 5.1 & 0.2 & 0.72 & 1.0 \\
353, Whole Sky & 1246 & 13.9 & -1.9 & 0.28 \\
\hline
217, Zone I & 998 & 12.9 & 1.6 & 0.19 & 98.0 \\
217, Zone II & 10 & 12.8 & -0.1 & 0.19 & 1.0 \\
217, Zone III & 5 & 9.9 & 1.4 & 0.26 & 1.0 \\
217, Whole Sky & 1013 & 12.9 & 1.6 & 0.19 \\
\hline
143, Zone I & 998 & 11.2 & -1.1 & 0.23 & 99.0 \\
143, Zone II & 8 & 10.4 & 1.8 & 0.34 & 1.0 \\
143, Whole Sky & 1006 & 11.2 & -1.1 & 0.23 \\
\hline
100 (HFI), Zone I & 998 & 9.9 & -3.5 & 0.36 & 99.0 \\
100 (HFI), Zone II & 9 & 9.6 & -2.6 & 0.38 & 1.0  \\
100 (HFI), Whole Sky & 1007 & 9.9 & -3.5 & 0.36 \\
\hline
100 (LFI), Zone I & 1073 & 14.2 & -3.9 & 0.31 & 99.0 \\
100 (LFI), Zone II & 10 & 14.2 & -4.0 & 0.31 & 1.0 \\
100 (LFI), Whole Sky & 1083 & 14.2 & -3.9 & 0.31 \\
\hline
70, Zone I & 748 & 13.6 & -6.4 & 0.37 & 99.0 \\
70, Zone II & 8 & 14.5 & -7.5 & 0.37 & 1.0 \\
70, Whole Sky & 756 & 13.6 & -6.4 & 0.37 \\
\hline
44, Zone I & 873 & 16.2 & -9.6 & 0.44 & 99.0 \\
44, Zone II & 4 & 9.2 & -2.3 & 0.64 & 1.0 \\
44, Whole Sky & 877 & 16.1 & -9.5 & 0.44 \\
\hline
30, Zone I & 859 & 13.8 & -8.7 & 0.49 & 99.0 \\
30, Zone II & 4 & 21.7 & -19.5 & 0.77 & 1.0 \\
30, Whole Sky & 863 & 13.9 & -8.8 & 0.49 \\
\hline
\end{tabular}
\caption{This Table shows similar characteristics than
the ones in Table 4. The number of point sources in the
complete catalogue is indicated in the second column.
Absolute and real values of the amplitude estimation mean
errors are provided in columns 3 and 4. Column 5 shows the
minimum flux above which all point sources are detected.
The last column gives an estimation of the percentage of 
the sky covered by each zone.}
\end{center}
\end{minipage}
\end{table*}

\section{Point Source Populations: Spectral Index Estimation}
A multifrequency analysis allows us to find point sources in
coincidence at several frequencies. Assuming a certain frequency
dependence for the intensity we can determine spectral indices 
characterizing the different point source populations that will
be observed by Planck.

We model the frequency dependence of the point source intensity in
a simple way

\begin{equation}\label{eq:PSInten}
I = I_o{\bigg(\frac{\nu}{\nu_o}\bigg)}^{\alpha},
\end{equation}
being $\alpha$ the spectral index that will define the
different point source populations.

In Table 7 we give an estimation of the number of point
sources that are found in coincidence in different channels,
using the $5\sigma_w(R_{opt})$ criterion data. By fitting
the point source estimated amplitudes to the expression
given above we calculate the spectral indices given in
columns 6, 7 and 8. For comparison, the true values of
these spectral indices are given in columns 3, 4 and 5.
The mean errors are presented in the last column of Table 5.
For spectral indices close to zero (flat spectrum) we give
the absolute error, whereas for typical spectral indices of
infrared galaxies relative errors are provided.

As shown in Table 7, and in agreement with the input model,
two main source populations can be identified in our catalogue
of detected sources. Infrared selected sources -- starburst
and late type galaxies at intermediate to low redshift and
high redshift ellipticals -- with spectral indices close to
$\sim 2.5$ are detected at high frequency. On the other hand,
radio selected flat--spectrum AGNs (radio--loud quasars, blazers,
etc.) are the dominant population in the low frequency channels.
At intermediate frequencies (353, 217 and 143 GHz) the spectral
index is $\sim -0.5$. The dominant source population in these
frequencies (from the Toffolatti et al. 1998 model) is the infrared
one, and it is so flat because the point source model has
taken these frequencies as the ones where the dominant galaxy
emission is turning from the dust emission to the synchrotron one.
The choice of the turning frequency is highly model--dependent
because of the poor knowledge of the galaxy emission at this 
spectral range. Future observations (as the Planck one) will help
us to better determine the spectral behavior of point sources at
these frequencies. In particular, to discriminate between
flat--spectrum and infrared populations in these channels,
we need to know more information in the adjacent ones.

Some point sources can be followed through three or more adjacent
channels being the spectral index estimation quite good, with low
errors. At high as well as at low frequencies, the knowledge of 
the spectral behaviour can help to increase the number of point 
sources detected.

\begin{table*}
\begin{center}
\label{table:SpectralIndex}
\begin{tabular}{|c|c|c|c|c|c|c|c|c|}
\hline
Channels (GHz) &
$\sim N(>5\sigma_{w})$ &
$\alpha_{min}$ &
$\overline{\alpha}$ &
$\alpha_{max}$ &
${\alpha_{min}}^{est}$ &
${\overline{\alpha}}^{est}$ &
${\alpha_{max}}^{est}$ &
Mean Error\\
\hline
857 -- 545 & 2550 & 1.95 & 2.46 & 2.89 & 1.69 & 2.29 & 3.02 & 9.55\% \\
\hline
857 -- 545 -- 353 & 810 & 2.49 & 2.71 & 3.09 & 2.38 & 2.64 & 3.09 & 4.21\%\\
\hline
353 -- 217 -- 143 & 300 & -0.52 & -0.55 & -0.58 & -0.65 & -0.49 & -0.33 &
0.13\\
\hline
100 -- 70 & 375 & -0.17 & -0.16 & -0.15 & -0.18 & -0.04 & 0.12 & 0.13 \\
\hline
70 -- 44 & 375 & -0.12 & -0.10 & -0.08 & -0.36 & -0.28 & -0.16 & 0.18 \\
\hline
44 -- 30 & 300 & 0.07 & 0.08 & 0.09 & -0.05 & 0.24 & 0.53 & 0.29 \\
\hline
100 -- 70 -- 44 -- 30 & 300 & -0.07 & -0.06 & -0.05 & -0.18 & -0.14 &
-0.10 & 0.08\\
\hline
\end{tabular}
\caption{Spectral indices for point sources found in coincidence
at the frequencies indicated in column 1. An estimation of the
number of point sources found in coincidence in the whole sky
(outside the Galactic plane) is given in the second column. Third,
fourth and fifth columns give minimum, mean and maximum
values of the spectral indices as calculated from the original
point source simulation. The spectral index estimation obtained
from the MHW method is presented in columns 6, 7 and 8. The mean
error in the spectral index estimation (absolute error for
spectral index close to zero are given) is shown in the last column.}
\end{center}
\end{table*}

\section{Conclusions and Discussion}
The MHW method introduced in C00 to detect point sources in microwave
maps has been revised and improved in this paper. We are able to
establish the optimal Mexican Hat scale to maximize the detection
of point sources. Moreover, the amplitude estimation is now based
on a $\chi^2$ estimator that takes into account the correlation
between different scales. Using this method we have estimated the
number and amplitude of point sources that will be detected over
the whole sky (outside the Galactic plane) in any of the Planck
channels. We provide information about all the point sources to
be detected above a limiting flux as well as all the ones that
will be part of a complete catalogue. We considered two different
criterions to determine the point source catalogues. With the
first one, we select those sources which are above
$5\sigma_w(R_{opt})$  and have an optimal $\chi^2$. With the second
one, we select those sources which are above certain fluxes, 
such that the amplitude estimation error is $< 50\%$, allowing a
$5\%$ of spurious detections. Assuming the first criterion,
the amplitude estimation errors are in all channels below $15\%$,
whereas the second one provides a slightly worse estimation. On
the other hand, the flux limits obtained with the last one are
lower than the ones obtained with the first criterion (except for
the highest channels, where the number of detections with both
criterions are enough to determine the flux threshold).

It is important to notice that this
\emph{Planck Extragalactic Point Source Catalogue}
will be directly obtained from the observed maps, without
denoising them or separating the foregrounds. No assumption
is needed about the nature of the signal and noise that will
be present in the observed maps. The only assumption made in
this work is the Gaussian and white nature of the instrumental
noise. We have not tested how the presence of other kinds of
noise can modify the results. However, it is expected that
only correlated noise with a variation scale around the
point source one would be a handicap to our method.
In this case, a study with an optimal pseudofilter,
in the manner proposed by Sanz et al., 2000 should be
done. In that work the MHW detection method and the
pseudofilter one are compared in presence of different kinds
of noise. The results demonstrate that there are only small
differences in the amplification achieved with the two filters.

The study here presented is done considering all the Planck channels.
This allows us to look for point sources coincident in different
frequency maps. A simple spectral dependence model is assumed and
spectral indices are estimated for different populations present
in the extracted \emph{Planck Extragalactic Point Source Catalogue}.
As discussed in the previous Section (and shown in Table 7),
the proposed detection method is able to recognize the two main
source populations present in the simulated maps as well as to give
spectral index estimations with errors below $10\%$. Moreover,
knowledge of spectral behaviour can be used to detect more point sources.

The MHW has proved to be the most powerful tool to extract point
sources from microwave maps. Using information at different scales
provides estimations of point source amplitudes with very small
errors. The MHW method has advantages over the other methods also
applied to microwave maps as explained in the Introduction. We have
also done the exercise of comparing the MHW with the image analysis
package SExtractor. We have analysed a square
$12.8^{\circ}\times 12.8^{\circ}$ patch at 857 GHz.
An especially optimal case for point source estimation has been
chosen: a very low and homogeneous dust emission patch
(belonging to Zone I with rms = $5.99\times 10^{-2}$, close
to the lowest limit in this zone. See Table 2). We have studied
different ways to detect and subtract point sources using SExtractor.
The best results have been obtained performing a background
estimation with a mesh of $13.5^{'}\times13.5^{'}$. After the
background subtraction, the map has been convolved with the
optimal MHW provided by SExtractor. We have checked that a
gaussian filter works worse than the MHW. The results, for
the minimum flux reached in our method (applying first criterion),
are: 74 point sources detected using SExtractor, with a mean error
of $29.02\%$ and 68 ones detected using our technique, with a mean
error of $8.83\%$. SExtractor can detect 6 point sources more than
our method, but these detections have a large error. In fact,
a high number of the SExtractor detections have an error $> 100\%$,
especially for low amplitude point sources. We reject this spurious
detections in our method because of the scale fit used. We have also
compared the results for the completeness flux. In this case, we
can reach a slightly lower flux than SExtractor, and the results
are: 35 point sources detected using SExtractor, with a mean error of
$14.41\%$ and 43 ones using our method, with a mean error of
$4.95\%$. We want to remark that the map analysed is the most suitable
case for the SExtractor performance. The power of our method is
not only supported by the choice of the MHW as filter, but also
by the study of the optimal scale and the scale fit. This allows
us to optimize both the detection and the amplitude estimation,
as well as to reject spurious detections.

The MHW method works better at high frequencies. Dust is
dominating the maps at those frequencies, having a
\emph{coherence scale} greater than the antenna beam
(point source scale). However, at low frequencies the main
background is CMB, whose \emph{coherence scale} is similar or
lower than the antenna dispersion. In this situation both point
sources and background have a similar variation scale. We have
proved that an optimal selection of the MHW scale is important
in order to reach high amplifications. We must study each image
separately and, in particular, we need to compute the power
spectrum from each image in order to calculate the wavelet
coefficient dispersion for several MHW scales and look for
the maximum amplification.

In those cases where the background has a similar variation scale
than the point sources or it has a high emission, the MHW tends
to an optimal scale lower than the antenna dispersion. On the
other hand, on those patches having a background with a variation
scale greater than the point sources (usually dust emission at
high frequencies) we are able to detect a large number of point
sources and we can reach very low levels; in this case the
analysing scale of the MHW should be greater than the antenna
dispersion. At this point, the noise (with pixel scale variation)
plays a very important role and makes the MHW take this
relatively large optimal scale. Therefore, given a particular
image, the optimal MHW scale arises from a compromise between
the intensity and \emph{coherence scale} of the dominant
foreground (or CMB background) and the noise amplitude.

One could think in different ways of improving the number of
detections. Denoising the maps before applying the detection method
could be important to increase the number of point sources
detected in those maps where the noise is the main problem:
in the high frequency channels. The denoising method should
be such that it preserves the Gaussian shape of point sources
in order to preserve the efficiency of the MHW.

The antenna size also plays an important role since it is
limiting the number of point sources detected at frequencies
at which the CMB is the dominating signal. Only a better angular
resolution at low frequencies (below the CMB \emph{coherence scale})
could improve the number of detections. In this case (and also in
the previous one) we can increase the number of detections using
the information about the spectral index of the point source
families we have detected. Another way to improve both the number
of detections and the amplitude estimation would be to combine the
MHW method with the MEM one. Since the last one has problems in
subtracting the brightest sources in the maps, the MHW would
represent a natural complement to MEM. A collaborative effort in
this direction is being carried out at present (Vielva et al. 2000).

A generalization of the method to account for possible elliptical
asymmetries in the gaussian beam profile can be easily done by
modifying the isotropic MHW to have the same asymmetry. However,
in the specific case of the Planck mission an important effort is
presently being done in order to avoid beam asymmetries.
Otherwise, due to the scanning strategy which will cover the
entire sky with circles, small asymmetries would imply a strong
degradation of the data.

In the near future we plan to implement the MHW method to work
directly on maps covering the whole sphere, using the pixelization
adopted by Planck. All sky point source simulations will clearly
allow for the presence of brighter sources than considered in
this work. These sources will be easily detected by the MHW method.

Finally we would like to emphasize that the MHW method will allow
for a direct extraction of the
\emph{Planck Extragalactic Point Source Catalogue} before any
foreground removal or separation is done. The catalogue will be
obtained with amplitude estimation errors below $15\%$. This will
be of great value, not only increasing the number of extragalactic
sources known at Planck frequencies but also providing information
about the spectral behaviour of different populations, among which
there is an important lack of knowledge at present.

\section*{Acknowledgments}
We thank B. Guiderdoni for kindly providing us with the
source counts predicted by his model E. 
We also thank B. T. Draine and A. Lazarian for providing
us with the emissivity predicted by their spinning dust model.
We thank Julio
Gallegos for his help with synchrotron emission simulations
and Luis Tenorio and Diego Herranz for useful comments.
We also thank R. F. Bouchet for helpful suggestions and
comments. PV acknowledges support from Universidad de Cantabria
fellowship as well as the CfPA hospitality during November 1999.
JMD acknowledges support from a Spanish MEC fellowship
FP96 20194004. We thank the Comisi{\'o}n Conjunta
Hispano--Norteamericana de Cooperaci{\'o}n Cient{\'\i}fica y
Tecnol{\'o}gica ref. 98138, Spanish DGESIC Project no.
PB98-0531-c02-01 for partial financial support.
EMG, LC and JLS thank FEDER Project no. 1FD97-1769-c04-01
for partial financial support. EMG, LC, JMD and JLS thank
the EEC Project INTAS-OPEN-97-1992 for partial financial support.

\bsp

\label{lastpage}


\begin{thebibliography}{99}
\bibitem{b1} {Aghanim, N. \& Forni, 0., 1999, A\&A, 347, 409.}
\bibitem{b2} {Baccigalupi, C., Bedini, L., Burigana, C., De Zotti, G.,
Farusi, A., Maino, D., Maris, M., Perrota, F., Salerno, E.,
Toffolatti, L., Tonazzini, A., 2000, astro-ph/0002257} 
\bibitem{b3} {Bouchet, F. R., Gispert, R. \& Puget, J. L., 1996
in Drew, E., ed., Proc. AIP Conf. 384, The mm/sub-mm foregorunds
and future CMB space missions. AIP Press, New York, p. 255}  
\bibitem{b4} {Bouchet, F. R., Prunet, S. \& Sethi, S. K., 1999,
MNRAS, 302, 663}
\bibitem{b5} {Boulanger, F. \& P{\'e}rault, M., 1988, ApJ, 330, 964.}
\bibitem{b6} {Cay{\'o}n, L., Sanz, J. L., Barreiro, R. B.,
Mart{\'\i}nez-Gonz{\'a}lez, E., Vielva, P., Toffolatti, L., Silk, J.,
Diego, J. M. \& Arg{\"u}eso, F., 2000, C00, MNRAS, 315, 757.}
\bibitem{b7} {Diego, J. M., Mart{\'\i}nez-Gonz{\'a}lez, E., Sanz, J. L.,
Cay{\'o}n, L. \& Silk, J., 2000, MNRAS submitted.}
\bibitem{b8} {Draine, B. T. \& Lazarian, A., 1998, ApJ, 494, L19.}
\bibitem{b9} {Ferreira, P, Magueijo, J. \& Silk, J., 1997,
Phys. Rev. D, 56, 4592.}
\bibitem{b10} {Finkbeiner, D. P., Davis, M. \& Schlegel, D. J., 1999,
ApJ, 524, 867}
\bibitem{b11} {Guiderdoni, B., Hivon, E., Bouchet, F. R.
\& Maffei, B., 1998, MNRAS, 295, 877}
\bibitem{b12} {Haslam, C. G. T., Salter, C. J., Stoffel, H. \& Wilson,
W. E., 1982, A\&AS, 47, 1}
\bibitem{b13} {Hobson, M. P., Jones, A. W. Lasenby, A. N. \& Bouchet,
F. R., 1998, MNRAS, 300, 1}
\bibitem{b14} {Hobson, M. P., Barreiro. R. B., Toffolatti, L.,
Lasenby, A. N., Sanz, J. L., Jones, A. W. \& Bouchet F. R.,
1999, MNRAS. 306, 232.}
\bibitem{b15} {Jonas, J. L., Baart, E. E. \& Nicolson, G. D., 1998,
MNRAS, 297, 977}
\bibitem{b16} {Mandolesi, N. et al. 1998: Proposal submitted to ESA
for the Planck Low Frequency Instrument.}
\bibitem{b17} {Puget, J. L. et al. 1998: Proposal submitted to ESA
for the Planck High Frequency Instrument.}
\bibitem{b18} {Reich, P. \& Reich, W., 1986, A\&AS, 63, 205}
\bibitem{b19} {Sanz, J. L., Arg{\"u}eso, F., Cay{\'o}n, L., 
Mart{\'\i}nez-Gonz{\'a}lez, E., Barreiro, R. B. \& Toffolatti, L.,
1999a, MNRAS, 309, 672.}
\bibitem{b20} {Sanz, J. L., Barreiro, R. B., Cay{\'o}n, L.,
Mart{\'\i}nez-Gonz{\'a}lez, E., Ruiz, G. A., D{\'\i}az, F. J., Arg{\"u}eso, F.,
Silk, J. \& Toffolatti, L., 1999b, A\&AS, 140, 99.}
\bibitem{b21} {Sanz, J. L., Herranz, D. \& Mart{\'\i}nez--Gonz{\'a}lez, E.,
2000, ApJ, accepted.}
\bibitem{b22} {Seljak, U. \& Zaldarriaga, M., 1996, ApJ, 469, 437.}
\bibitem{b23} {Tegmark, M. \& Oliveira-Costa, A., 1998, TOC98,
ApJ, 500, 83.}
\bibitem{b24} {Tenorio, L., Jaffe, A. H., Hanany, S. \&
Lineweaver, C. H., 1999, MNRAS, 310, 823.}
\bibitem{b25} {Toffolatti, L., Arg{\"u}eso, F., De Zoti, G.,
Mazzei, P., Franceschini, A., Danese, L. \& Burigana, C.,
1998, MNRAS, 297, 117.}
\bibitem{b26} {Vielva, P., Barreiro, R. B., Hobson, M. P.,
Mart{\'\i}nez--Gonz{\'a}lez, E., Lasenby, A. N., Sanz, J. L.
\& Toffolatti, L., 2000, MNRAS, submitted.}

\end{thebibliography}
\end{document}